\documentclass[aps,showpacs,epsfig,twocolumn]{revtex4}
\usepackage{epsfig}

\newcommand{\be}{\begin{equation}}
\newcommand{\ee}{\end{equation}}
\newcommand{\bea}{\begin{eqnarray}}
\newcommand{\eea}{\end{eqnarray}}
\usepackage{color}
\begin{document}

\title{\bf\Large{Applied electric field instead of pressure in H-based superconductors.}}

\author{Naoum Karchev}

\affiliation{Department of Physics, University of Sofia, 1126 Sofia, Bulgaria }

\begin{abstract}
In our desire to give a new suggestion for H-based superconductors experiments we present a theoretical framework for understanding the impact of an applied electric field on pressured hydride superconductors. We study a material at pressure $p$, when it possesses insulator-superconductor transition, at the respective superconducting critical temperature $T_{cr}$. The theory shows the applied electric field penetrates the material and forces the Cooper pairs to Bose condensate. If one applies an electric field and then increases the temperature, the theory predicts novel critical temperature $T^{el}_{cr}$  higher than $T_{cr}$. Therefore, the system has a higher superconducting critical temperature if we apply an electric field instead of increasing the pressure.
The result shows that in the case of carbonaceous sulfur hydride at $234Gpa$ and near but below critical temperature $T_c=283K$, applying a sufficiently strong electric field, we can bring the superconducting critical temperature close to 300K. 
\end{abstract}


\maketitle

\section {\bf Introduction}

The present article is a sequel to article \cite{Karchev17}, where the system of equations which describes the electrodynamics of s-wave superconductors is derived, and articles \cite{Karchev17b,Karchev19}, where results of numerical calculations are presented. We use our knowledge to look at a really  existing systems, H-based superconductors, to show that the effect of pressure can be obtained by applying an electric field.

The theoretical predictions are important, since they allow to focus experimental research on specific chemical compositions and also to suggest appropriate conditions for the synthesis.	
In his theoretical studies Ashcroft predicted that metalized hydrogen \cite{Ashcroft68} or hydrogen rich alloys \cite{Ashcroft04} can possess high temperature superconductivity. Ashcroft's idea is that 
the hydrogen is the lightest element, and if it can be compressed in a solid state, it could become superconductor at a very high transition temperature $T_c$ due to the strong electron-phonon coupling.
Accurate electronic structure and electron-phonon coupling calculations predicted high $T_c$ for metallic hydrogen \cite{Gross10,McMahon11,Duan15,Pickard15}. Thanks the development of high-pressure techniques numerous experiments 
discussed the prediction. An important discovery leading to room-temperature superconductivity is the pressure-driven hydrogen sulfide with a confirmed transition temperature of 203 K at 155 GPa \cite{Drozdov15}. The most recent examples of a metal hydride are lanthanum hydride which has $Tc = 250 - 260 K$ at $ 180 - 200 GPa $ \cite{Drozdov19,Somayazulu19,Hong20} and carbonaceous sulfur hydride with room-temperature $T_c =287.7 K$ achieved at $267GPa$ \cite{Snider20}. More than ten hydrogen-rich compounds under high pressure have been found to be high temperature conventional superconductors: yttrium superhydride \cite{Troyan21,Snider21}, thorium hydride \cite{Semenok20}, praseodymium superhydride \cite{Zhou20}, barium superhydride \cite{Chen21} and others \cite{Sanna16,Gorkov18,Meng19,RevHSc20,Pickard20}.

The earliest study of the electrodynamics of s-wave superconductors is attributed to London brothers \cite{London35}. They supplemented the Maxwell system of equations with set of equations to explain the electrodynamics of superconductors and more particularly the Meissner-Ochsenfeld effect. The quantum-mechanical foundation of these equations was discussed in phenomenological Ginzburg-Landau theory \cite{GL50}. To extend this work, to include the electric field,
we start with field-theoretical action of the time-dependent Ginsburg-Landau (TDGL) theory \cite{Gorkov68, Thompson70, Tinkham75}.
\bea\label{HSE1}
S & = & \int d^4x\left [-\frac 14 \left (\partial_{\lambda}A_{\nu}-\partial_{\nu}A_{\lambda}\right)\left (\partial^{\lambda}A^{\nu}-\partial^{\nu}A^{\lambda}\right)\right. \nonumber  \\  
& + & \left. \frac {1}{D}\psi^*\left (i\partial_{t}-e^*\varphi\right)\psi \right. \\
& - & \left.  \frac {1}{2m^*} \left (\partial_{k}-ie^*A_{k}\right)\psi^* \left (\partial_{k}+ie^*A_{k}\right)\psi \right. \nonumber \\
& + & \left.  \alpha \psi^*\psi -\frac {g}{2}\left(\psi^*\psi \right)^2\ \nonumber \right],
\eea
written in terms of gauge four-vector electromagnetic potential $"A"$ and complex scalar field $"\psi"$-the superconducting order parameter. We use the notations $x=(x^0,x^1,x^2,x^3)=(\upsilon t,x,y,z)$, $\upsilon^{-2}=\mu\varepsilon$, where $\mu$ is the magnetic permeability and $\varepsilon$ is the electric permittivity of the superconductor.  The constant $D$ is the normal-state diffusion, ($e^*,m^*$) are effective charge and mass of superconducting quasi-particles and $\varphi=\upsilon A_0$ is the electric scalar potential. The index $k$ runs $k=x,y,z$ and $\alpha$ is a function of the temperature $T$
\begin{equation}\label{HSE2}
\alpha=\alpha_0(T_c-T),\end{equation}
where $T_c$ is the superconducting critical temperature and $\alpha_0$ is a positive constant. In superconducting phase $\alpha$ is positive. 
 
The action (\ref{HSE1}) is invariant under $U(1)$ gauge transformations. We represent the order parameter in the form $\psi(x) =  \rho(x)\exp {[ie^*\theta(x)]}$,
where $\rho(x)=|\psi(x)|$ is the gauge invariant local density of Cooper pairs and introduce the gauge invariant vector  $ Q_{k}\, = \,\partial_{k}\theta+ A_{k}$, scalar $Q = \partial_{t}\theta+ \varphi$ and antisymmetric  $F_{\lambda\nu}\, = \,\partial_{\lambda}A_{\nu}-\partial_{\nu}A_{\lambda}$ fields.

 The system of equations which describes the electrodynamics of s-wave superconductors is \cite{Karchev17}:
\bea\label{HSE111}
& & \overrightarrow{\nabla}\times\textbf{B}\,=\,\mu\varepsilon\frac {\partial \textbf{E}}{\partial t}-\frac {e^{*2}}{m^*}\rho^2\textbf{Q} \label{HSC3} \nonumber\\
& & \overrightarrow{\nabla}\times\textbf{Q}\,=\,\textbf{B}\label{MSc102}\\
& & \overrightarrow{\nabla}\cdot\textbf{E}\,=\,\frac {\mu\varepsilon e^{*}}{D}\rho^2 \label{EScCp13} \nonumber\\
& & \overrightarrow{\nabla} Q+\frac {\partial \textbf{Q}}{\partial t}\,=\,-\textbf{E}\label{MSc104} \nonumber \\
& & \frac {1}{2m^*}\Delta\rho +\alpha \rho-g\rho^3- \frac {e^{*}}{D} \rho Q-\frac {e^{*2}}{2m^*}\rho \textbf{Q}^2=0.\nonumber
\eea
where the electric $\bf E$ and magnetic $\bf B$ fields are introduced in a standard way,
$(F_{01},F_{02},F_{03})=\textbf{E}/\upsilon$, $(F_{32},F_{13},F_{21})=\textbf{B}$.

\section{The impact of the applied electric field on the superconductivity} 
We are interested in the system of equations for time-independent fields without magnetic field $\textbf {B}=\textbf{Q}=0$.
In the case of superconductor with slab geometry, the fields depend on one of the coordinates "z"  and the electric field vector has one nonzero component $\textbf{E}=(0,0,E)$.
In normal phase $\alpha$ is negative. 

We introduce dimensionless functions $f_1(\zeta)=-Q(\zeta)\sqrt{2m^*|\alpha|}/E_0$, $f_2(\zeta)=E(\zeta)/E_0$ and $f_3(\zeta)=\rho(\zeta)\sqrt {g/|\alpha|}$ of a dimensionless distance $\zeta=z\sqrt {2m^*|\alpha|}$, where  $\textbf{E}_0=(0,0,E_0)$ is the applied electric field. One can write the system of equations  by means of the dimensionless functions 
\begin{eqnarray}\label{HSE9}
& & \frac{df_1}{d\zeta}\,=\,f_2  \nonumber \\
& & \frac{df_2}{d\zeta}\,=\frac {\beta}{\gamma}f_3^2  \\
& & \frac {d^2f_3}{d^2\zeta}\,-\,f_3\,-\,gf_3^3\,+\,\gamma f_1 f_3 =0. \nonumber
\end{eqnarray}
The solutions of the system (\ref{HSE9}) depend on the parameters $\beta$ and $\gamma$ 
\begin{eqnarray}
& & \beta\,=\, \frac {e^{*2}\mu\varepsilon}{2m^* D^2|\alpha|g}\label{HSE10} \\
& & \gamma\,=\,\frac {e^*}{D\sqrt{2m^*|\alpha|^3}} E_0 \label{HSE11}. 
\end{eqnarray}
When the temperature is fixed, $\alpha$ (\ref{HSE2}) is fixed, the parameter  $\gamma$ (\ref{HSE11})
varies following the variation of applied electric field $E_0$. 
We solve the system of equations (\ref{HSE9}) for different values of $\gamma$ and fixed $\beta$.

\begin{figure}[!ht]
\epsfxsize=\linewidth
\epsfbox{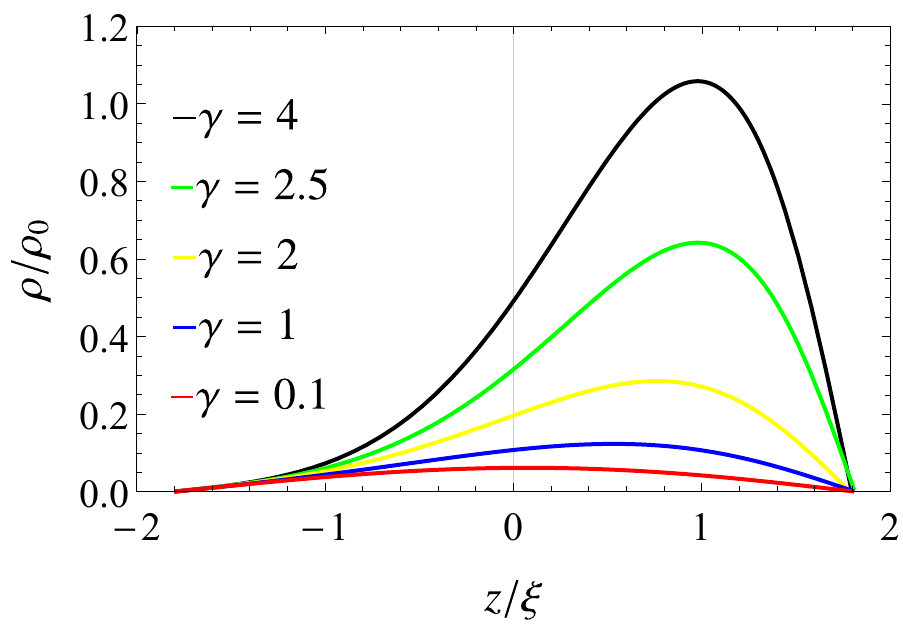} \caption{\,\,{\bf The dimensionless function $\rho/\rho_0$ of a dimensionless distance $\zeta=z/\xi$,} where $\rho_0=\sqrt{|\alpha|/g}$ and $\xi=1/\sqrt{2m^*|\alpha|}$, is plotted. Increasing the applied electric field, keeping the parameter $\beta$ fixed, $\beta=2.5$, one increases the parameter $\gamma$: $\gamma=0.1$-red line, $\gamma=1$-blue line,  $\gamma=2$-yellow line, $\gamma=2.5$-green line, $\gamma=4$-black line.  Figures show that applying electric field the Cooper pairs Bose condense $\rho/\rho_0>0$ and the onset of superconductivity starts while the temperature is higher then critical one $\alpha<0$. Below $\gamma_{cr}=0.09$ there is no superconductivity }\label{fig1-MScIII}
\end{figure}

The dimensionless function $f_3(\zeta)$ of a dimensionless distance $\zeta$ is depicted in figure (\ref{fig1-MScIII}) for different values of $\gamma$ proportional to applied electric field and $\beta$ is fixed $\beta = 2.5$. For slab geometry $\zeta$ runs the interval $-1.8\leq \zeta \leq 1.8$. Increasing the applied electric field one increases the parameter $\gamma$. Figures show that applying electric field the Cooper pairs Bose condense $\rho >0$ and the onset of superconductivity starts while the temperature is higher then critical one $\alpha<0$, when the electric field is off. The figures also show that decreasing the electric field, i.e. the gamma parameter, the density of Cooper pairs decrease. There is a critical value  $\gamma_{cr}$ below which there is no Bose condensation of the Cooper pairs and respectively no superconductivity. Accordingly there is critical value $\alpha_{cr}$  and one can calculate it from equation (\ref{HSE11}).
We represent $\alpha_{cr}$ using equation (\ref{HSE2}) $|\alpha_{cr}|=\alpha_0 (T_{cr}^{el}-T_c)$, where $T_{cr}^{el}$ is the superconducting transition temperature in the presence of electric field. Straightforward calculations lead to a formula for $T_{cr}^{el}$ as a function of applied electric field  
\be\label{HSE12}
T_{cr}^{el} - T_{cr}\,=\,\frac {1}{\alpha_0}\left(\frac {e^*}{D\gamma_{cr}\sqrt{2m^*}}\right)^{\frac 23} E_0^{\frac 23}\ee

The critical value of $\gamma$ parameter weakly depends on $\beta$ parameter. The last one depends on $g$ constant in the Ginzburg-Landau theory (\ref{HSE1}), which is characteristic of material. In the case of metalized hydrogen or hydrogen rich alloys under pressure $g$ and respectively $\beta$ parameters depend on the pressure, therefore $\gamma_{cr}$ weakly depends on the pressure.  

The critical value of $\gamma$ depends of the size of the sample along z axis.

In the case of slab geometry, the density of the Bose-condensed Cooper pairs $\rho (z)$ depends on the coordinate $z$. This inhomogeneity is a function of the applied electric field. Increasing $\gamma $ increases the density $\rho$ and the maximum moves to one of the slab surfaces. This can be interpreted as a projection of the superconductivity on the slab surface.

\section{The impact of the applied electric field on the H-base superconductors} 
\begin{figure}[!ht]
	\epsfxsize=\linewidth
	\epsfbox{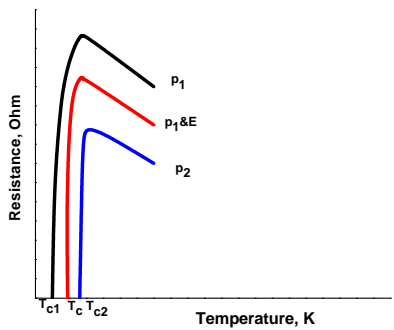} 
	\vskip -1.5 cm
	\caption{{\bf We consider pressured hydride superconductor at pressure within interval (p1; p2) where the system possesses insulator-superconductor transition.} The typical resistance-temperature curves are depicted, where Tc1 and Tc2 are the respective critical temperatures. 
    In particular for sulfur hydride \cite{Drozdov15} $p_1=107GPa, p_2=122GPa$ and $T_{c1}=10K, T_{c2}=20K$, for  carbonaceous sulfur hydride $p_{c1}=234 GPa$ and $T_{c1}=283 K$ \cite{Snider20}. The red line in between is a hipotetic curve when electric field is applied and the critical temperature increases from $T_{c1}$ to $T_{c}$, as follows from equation (\ref{HSE12})}\label{(fig1)H-Sc-E}
\end{figure}

To apply our result in the case of H-based superconductors,  we consider pressured hydride superconductor at pressure within interval $(p_1,p_2)$ where the system possesses insulator-superconductor transition. The typical resistance-temperature 
curves are depicted in figure (\ref{(fig1)H-Sc-E}), where $T_{c1}$ and $T_{c2}$ are the respective critical temperatures. In particular for sulfur hydride \cite{Drozdov15} $p_1=107GPa, p_2=122GPa$ and $T_{c1}=10K, T_{c2}=20K$, for  carbonaceous sulfur hydride $p_{c1}=234 GPa$ and $T_{c1}=283 K$ \cite{Snider20}. 

The applied electric field penetrates the superconductor when $p=p_1$ and $T<T_{c1}$. Increasing the temperature we see that the superconductivity remains a state of the system at $T> T_{c1}$. Formula (\ref{HSE12}) shows that even a weak applied electric field leads to increase in the critical temperature of the superconductor. So applying an electric field we can achieve the same increase of the critical temperature $T_c^{el} > T_{c1}$ as increasing the pressure (Fig.\ref{(fig1)H-Sc-E}). 

The difficulty in equation (\ref{HSE12})  is the constant  in front of $E_0^{\frac 23}$. It involves so many parameters that it is impossible to calculate it theoretically. So the first few measurements should lead to the determination of the constant after which we can write the equation  (\ref{HSE12}) for a given material. 

The result shows that in the case of carbonaceous sulfur hydride \cite{Snider20} at $234Gpa$ and near but below critical temperature $T_c=283K$, applying a sufficiently strong electric field, we can bring the superconducting critical temperature close to 300K. 

Another phenomenon worth noting is when the system has Cooper pairs above the critical temperature. The applied electric field, at a temperature above the critical, Bose condenses the existing Cooper pairs and onsets the superconductivity. We can repeat the procedure at higher temperature and in this way to arrived at temperature above which there are no Cooper pairs. 
Our investigation designs an experimental way for verification of the pairing of Fermions prior the superconductivity, and to measure the new characteristic temperature  $T^*$ of the system, below which the formation of Cooper pairs begins. In the BCS theory $T_c=T^*$, but in the case of strong coupling superconductors $T^*>T_c$. 

Pairing without superconductivity in some superconducting semiconductors at low carrier concentration is discussed in \cite{Eagles69}. The fluctuations of Cooper pairs above $T_c$ lead to charge transfer. This so-called paraconductivity is discussed in \cite{Ullah91, Chen98, Larkin02}. Their contribution to the thermal transport in the normal state is theoretically analyzed in \cite{Iddo02}.

The main result in the present article is that an applied electric field forces the Cooper pairs to Bose condense, which in turn leads to increasing the superconductor critical temperature (Eq.\ref{HSE12}). It is important to make difference between electrons and electron pairs in electric field. The electrons expel the electric field, while the Cooper pairs Bose condense. This is why we consider materials with insulator-superconductor transition. On the other hand in metal superconductors the density of normal quasiparticles decreases at low temperature, because of the formation of Cooper pairs. At very low temperature there is no quasiparticle current, and the system of equations (\ref{HSE111}) is appropriate to describe the electrodynamics of s-wave superconductivity in these materials. 

In the present article, we consider a superconductor located between the plates of a capacitor. A completely different setting is used in \cite{Glover60}. In the process of charging a parallel plate condenser, electrons are added to one plane and removed from the other one. Using a superconducting film $70 A$ thick as one of the condenser plate it is possible to add to or subtract from the superconductor electrons. A field of $2.6\times10^5$ v/cm  produces a change of transition temperature on the order of $10^{-4} K$. The  electrostatic-field effect technique, which allows the carrier density in a given sample to be changed, is used to modulate the superconducting transition temperature in high $T_c$ superconductors \cite{Ahn03,Matthey07} and $(a-Bi)$ films \cite{Parendo05}.

An alternative theory is developed in \cite{Lipavsky06}.  
The authors investigated a superconducting phase transition in a thin metal layer with an electric field  $ E $ applied perpendicular to the layer. Following de Gennes' approach, 
the field $ E $ enters the boundary conditions for the Ginzburg Landau wave function.

The main goal of the article is the technical results, obtained in previous articles \cite{Karchev17,Karchev17b,Karchev19}, to reach experimental physicists.
A new experiment is proposed to discuss H-based superconductors.

Data availability 

The system of equations which describes the electrodynamics of s-wave superconductors (3) is derived by the author.
The details are available from the author on request.


\begin{thebibliography}{99}	

\bibitem{Karchev17} Naoum Karchev, Condens. Matter {\bf 2}, 20 (2017) (arXiv:1512.04284).
\bibitem{Karchev17b} Naoum Karchev and Tsvetan Vetsov, Condensed Matter {\bf 2} 31 (2017).
\bibitem{Karchev19} N.Karchev and T.Vetsov, International Journal of Modern Physics {\bf B 33}, 1950384 (2019). 
\bibitem{Ashcroft68} N. W. Ashcroft, Phys. Rev. Lett. {\bf 21}, 1748 (1968).
\bibitem{Ashcroft04} N. W. Ashcroft, Phys. Rev. Lett. {\bf 92}, 187002 (2004).
\bibitem{Gross10} P. Cudazzo, G. Profeta, A. Sanna, A. Floris, A. Continenza, S. Massidda, and E. K. U. Gross, Phys. Rev. {\bf B 81}, 134506 {2010}.
\bibitem{McMahon11} J. M. McMahon and D. M. Ceperley, Phys. Rev. {\bf B 84}, 144515 (2011).     
\bibitem{Duan15} D.Duan, X. Huang, F. Tian, D. Li,H.Yu, Y. Liu, Y. Ma,B. Liu, and T. Cui, Phys. Rev. {\bf B 91}, 180502R (2015).
\bibitem{Pickard15} I. Errea, M. Calandra, C. J. Pickard, J. Nelson, R. J. Needs, Y. Li, H. Liu, Y. Zhang, Y. Ma, and F. Mauri, Phys. Rev. Lett. {\bf 114}, 157004 (2015). 
\bibitem{Drozdov15} A. P. Drozdov, M. I. Eremets, I. A. Troyan, V. Ksenofontov, and S. I Shylin, Nature  {\bf 525}, 73 (2015). 
\bibitem{Drozdov19} A. P. Drozdov, P. P. Kong, V. S. Minkov, S. P. Besedin, M. A. Kuzovnikov, S. Mozaffari, L. Balicas, F.F. Balakirev, D. E. Graf, V. B. Prakapenka, E.
Greenberg, D. A. Knyazev, M. Tkacz, M.I. Eremets,  Nature {\bf 569}, 528 (2019).  
\bibitem{Hong20} F. Hong et al., Chin. Phys. Lett. {\bf 37}, 107401 (2020). 
\bibitem{Somayazulu19}  Maddury Somayazulu, Muhtar Ahart, Ajay K. Mishra, Zachary M. Geballe, Maria Baldini,
Yue Meng, Viktor V. Struzhkin, and Russell J. Hemley, Phys. Rev. Lett. {\bf 122}, 027001 (2019).   
\bibitem{Snider20} Elliot Snider, Nathan Dasenbrock-Gammon, Raymond McBride, Mathew Debessai, Hiranya Vindana, Kevin Vencatasamy, Keith V. Lawler, Ashkan Salamat, and Ranga P. Dias, Nature {\bf 586}, 373 (2020). 
\bibitem{Troyan21} Y. A. Troyan, et al., Adv. Mater. {2006832} (2021).   
\bibitem{Snider21} E. Snider et al., Phys. Rev. Lett {\bf 126}, 117003 (2021).   
\bibitem{Semenok20} D. V. Semenok et al.,  Materials Today {\bf 33}, (2020).    
\bibitem{Zhou20} D. Zhou et al., Science Advances {\bf 6}, eaax6849 (2020).
\bibitem{Chen21} W. Chen et al., Nature Comm {\bf 12} 273 (2021).  
\bibitem{Sanna16} José A. Flores-Livas, Antonio Sanna, and E. K. U. Gross, Eur.Phys. J. {\bf B 89}, 63 (2016).             
\bibitem{Gorkov18} L. P. Gor'kov, V. Z. Kresin, Colloquium: Rev. Modern Phys., {\bf 90}, 011001 (2018). 
\bibitem{Meng19} Yue Meng, Viktor V. Struzhkin, and Russell J. Hemley, Phys. Rev. Lett. {\bf 122}, 027001 (2019). 
\bibitem{RevHSc20}  José A. Flores-Livas, Lilia Boeri, Antonio Sanna, Gianni Profeta, Ryotaro Arita, Mikhail Eremets, Phys. Rep. {\bf 856}, 1 (2020).  
\bibitem{Pickard20} C.J. Pickard, I. Errea, M.I. Eremets, Annu. Rev. Condens. Matter Phys. {\bf 11}, 1 (2020). 
\bibitem{London35} F. London and H. London, Proc. R. Soc. London  {\bf Ser A 149}, 71 (1935).  
\bibitem{GL50}  V. L. Ginzburg and L. D. Landau, Zh. Eksp. Teor. Fiz. {\bf 20}, 1064 {1950}.  
\bibitem{Gorkov68}  L. P. Gor'kov and G. M. Eliashberg, Soviet Phys. JETP {\bf 27}, 328 (1968).  
\bibitem{Thompson70} R. S. Thompson, Phys. Rev. {\bf B 1}, 327 (1970).
\bibitem{Tinkham75} Michael Tinkham, { \emph{Introduction to Superconductivity}} (McGRAW-HIL Inc 1975).
\bibitem{Eagles69} D. M. Eagles, Phys. Rev {\bf 186},456 (1969).
\bibitem{Ullah91} Salman Ullah and Alan T. Dorsey, Phys. Rev. {\bf B 44},262 (1991).
\bibitem{Chen98} Qijin Chen, Ioan Kosztin, Boldizs\'{a}r Jank\'{o}, and K. Levin, Phys. Rev. Lett. {\bf 81}, 4708 (1998).
\bibitem{Larkin02} Anatoly Larkin and Andrei Varlamov, Chapter of \emph {Handbook on Superconductivity: Conventional and Unconventional Superconductors} edited by K.-H.Bennemann and J.B. Ketterson,  (Springer-Verlag, Berlin, 2002).
\bibitem{Iddo02} Iddo Ussishkin, S. L. Sondhi, and David A. Huse, Phys. Rev. Lett. {\bf 89}, 150405 (2002).
\bibitem{Glover60} R. E. Glover III and M. D. Sherril,  Phys. Rev. Lett. {\bf 5}, 248 (1960). 
\bibitem{Ahn03} C. H. Ahn, J.-M. Triscone, and J. Mannhart, Nature (London) {\bf 424}, 1015 (2003).
\bibitem{Matthey07} D. Matthey, N. Reyren , J.-M. Triscone and T. Schneider,  Phys. Rev. Lett. {\bf 98}, 057002 (2007).
\bibitem{Parendo05} Kevin A. Parendo, K. H. Sarwa, B.Tan, A. Bhattacharya, M. Eblen-Zayas, N. E. Staley, and A. M. Goldman, Phys. Rev. Lett. {\bf 94}, 197004 (2005). 
\bibitem{Lipavsky06} P. Lipavsk\'y, K. Morawetz, J. Koláček, and T. J. Yang, Phys. Rev. {\bf B 73}, 052505 (2006).
\bibitem[*]{byline} Electronic address: naoum@phys.uni-sofia.bg
\end{thebibliography}
\end{document}